# *Invitro* Pharmacological Evaluations Of Ethanolic Extract Of *Jatropha Maheshwari*

Sankar V[a], Anand Babu K*[b], Deepak M[c], Poojitha Mallapu[d], Raghu S[e], Anandharaj G[f]

[a]Associate professor, Department of Pharmacology, Srinivasan College of Pharmaceutical Sciences, Trichy- 621112, India

[b]Associate professor Department of Pharmaceutical Chemistry, GRTIPER, Tiruttani- 631209, India

[c]Research Scholar, Faculty of Pharmacy, SRIHER, Chennai-600116, India

[d]Assistant Professor, Department of Pharmacology, GRTIPER, Tiruttani- 631209, India

[d]Raghu S, Associate professor, Department of Pharmacology, Vinayaka Missions College of Pharmacy, Salem – 636008, India

[e]Anandharaj G, Vellalar College of Pharmacy, Thindal, Erode- 638012, India

**ABSTRACT:**

Objective:

To assess the antioxidant, wound healing, anti-ulcer, and anti-inflammatory properties of *Jatropha maheshwari.*

Methods:

*Jatropha maheshwari* was collected from Kanyakumari district and authenticated. Ethanol was used for continuous hot percolation extraction of the plant. Antioxidant activity was evaluated using DPPH and ABTS assays. The wound healing potential was assessed through a wound scratch assay using 3T3-L1 murine fibroblast cell lines. Anti-ulcer activity was measured using the acid-neutralizing capacity (ANC) and $H^+/K^+$-ATPase inhibitory activity methods. Anti-inflammatory effects were determined through dose-dependent studies of the ethanol extract.

**Results:**

**Antioxidant Activity**: The crude extract exhibited strong antioxidant capacity with percentage inhibition values of 88% (DPPH) and 75.7% (ABTS).

**Wound Healing Activity:** The wound closure rate in treated 3T3-L1 cell lines reached 97.88%, indicating potent wound healing properties.

**Anti-Ulcer Activity:** The extract demonstrated an 80.1% inhibition compared to the control when tested alongside omeprazole. ANC per gram of antacid was recorded at 20 and 26 for oral doses of 10 and 500 μg/ml, respectively.

**Anti-Inflammatory Activity:** Dose-dependent inhibition percentages of 30.1% (10 μg/ml) and 96.5% (500 μg/ml) were observed ($p<0.001$) relative to inflammation control. Additional percentages of 26.3% (10 μg/ml) and 42.6% (500 μg/ml) further confirmed the anti-inflammatory activity.

**Conclusion:**

*Jatropha maheshwari* demonstrates significant antioxidant, wound healing, anti-ulcer, and anti-inflammatory properties. Further research is warranted to elucidate the mechanisms underlying these pharmacological effects.

***Keywords:*** *Anti-ulcer, Anti-inflammatory, Ethanolic extract, Jatropha Maheswarii, Wound healing*

\* Corresponding author. Tel.: +91 8531010505.
*E-mail address:* anandanalyst85@gmail.com

## 1. Introduction:

The genus Jatropha belongs to the family, Euphorbiaceae and is distributed in India with 13 species. The name Jatropha is derived from the Greek words Jatros (doctor) andTrophe (nutrition) [1]. Many species belonging to this family have been used traditionally for their medicinal properties [2]. Among this, *Jatropha maheshwari* was chosen for the study. Among various species this genus, *Jatropha maheshwarii* Subr. and Nayar is an endemic species whose distribution is constrained to the southern coastal belts, plains and hilly regions of Kanyakumari, Thoothukudi and Tirunelveli districts of Tamil Nadu, extending to the west coast up to Thiruvananthapuram district of Kerala [3]. This plant is commonly called as 'Athalai,'Vel-athalai', and 'Kattamannaku' in Tamil. It is an evergreen under shrub, thick stem and dark green, petiole glabrous, leavesovate-lanceolate [4,5]. It is a fertile, drought hardy and rhizomatous plant which attains a height of about 2m having 22 chromosomes [6]. This plant is notable for its valuable traditional medicinal properties among the locals against rheumatism, eczema, ringworms and as an insecticide [7,8]. *Jatropha maheshwari* was scientifically evaluated for Antibacterial and antioxidant studies. In accordance with earlier research, *Jatropha maheswari* could have significant pharmacological activity. Hence the aim of the study was to find out the wound healing, antiinflammatory and anti-ulcer activity of *Jatropha maheshwari*.

**Figure 1:** *Jatropha Maheswari*

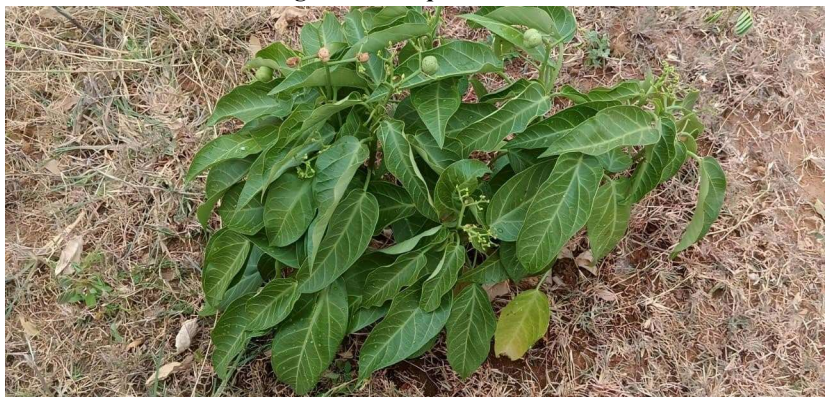

## 2. Collection and authentication:

The plant was selected according to Ethnobotanical survey and it was collected from kudankulam (Kanyakumari) Tamilnadu, India. The plant was authenticated by Dr.V.Nandhagopalan, Associate professor, PG research department of Botany in National college, Trichy, Tamilnadu, India. Authentication no is (*Jatropha maheshwarii* Subrm & M.P.Nayar - NCT/BOT/ PI/23/200046).

## 3. Preparation of Extract

Collected aerial plants were shade dried, dried aerial parts was pulverized by mortar and pestle. Grinded coarse material was passed sieve no (40). 500 gm packed in soxhlet apparatus for extraction by using ethanol as a solvent. The yield of the extract was found to be 28gm.

## 4. Materials Required

DMEM medium, Fetal Bovine Serum (FBS) and antibiotic solution were from Gibco (USA), 1X PBS was from Himedia, (India). 6 well tissue culture plate and wash beaker were from Tarson (India), ABTS, and DPPH from Merck. UV spectrophotometer (Perkin Elmer)

## 5. Invitro Antioxidant

### 5.1. ABTS Method:

The ABTS assay followed the method of Arnao et al. with modifications. Stock solutions included 7 mm ABTS and 2.4 mm potassium persulfate. The working solution was prepared by mixing these in equal quantities and incubating for 14 hours at room temperature in the dark. This solution was then diluted with 60 ml methanol to achieve an absorbance of $0.706 \pm 0.01$ units at 734 nm. Fresh ABTS solution was prepared for each assay. Various concentrations of MM extract (500, 250, 100, 50, and 10 µg/ml) were tested by mixing with 1 ml of the ABTS solution and measuring absorbance at 734 nm after 7 minutes. ABTS scavenging capacity was compared with ascorbic acid, and percentage inhibition was calculated as [(Abs control – Abs sample) / Abs control] × 100, where Abs control is the absorbance of ABTS radical in methanol and Abs sample is the absorbance of ABTS solution mixed with sample extract/standard. Triplicate measurements were performed for each determination (n = 3) [9].

### 5.2. DPPH Method:

Prepare a 0.1 mm DPPH solution in methanol. Mix 100 µl of this solution with 300 µl of sample MM at concentrations of 500, 250, 100, 50, and 10 µg/ml. Shake vigorously and let stand at room temperature for 30 minutes. Measure absorbance at 517 nm using a UV-VIS spectrophotometer, with ascorbic acid as the reference. Lower absorbance indicates higher radical scavenging activity. Calculate scavenging effect (%) using the formula: [(absorbance of control - absorbance of reaction mixture) / absorbance of control] × 100. Percentage inhibition and IC50 values were determined using GraphPad Prism 6.0 software.

## 6. *Invtiro* Wound Healing Procedure:

### 6.1. Cell Culture

3T3-L1 (Murine Fibroblast cells) cell line was purchased from National centre for cellular sciences, Pune, and were cultured in liquid medium (DMEM) supplemented 10% Fetal Bovine Serum (FBS), 100 µg/ml penicillin and 100 µg/ml streptomycin, and maintained under an atmosphere of 5% $CO_2$ at 37º C.

### 6.2. Wound Healing Assay

Wound healing assay was used to assess cell migration of both cancer and non-cancer cell lines upon treatments. 3T3-L1 cells were seeded into a six-well tissue culture dish and allowed to grow to 90% confluency in complete medium. Cell monolayers were wounded by a plastic tip (1 mm) that should be touched the plate. Wounded monolayers were then washed four times with medium to remove cell debris and incubated in 1 % FBS medium. The cells were treated with 433 µg/ml of MM sample and incubated for 48 h. Cells were monitored under an inverted microscope equipped with a camera. The wound area was measured using Image-J software (NIH, Bethesda, MD, USA). The wound area percentage was calculated as the wound area of control and treated sample [10, 11].

### 7. Invitro Antiulcer Activity:

#### 7.1. Assay of $H^+$-$K^+$ ATP ase [12, 13]

In the reaction mixture (40 mM Tris-HCl buffer, pH 7.4, including 2 mm MgC12 and 10 µg membrane protein) were incubated at various concentrations of the extract-EEJM (500, 250, 100, 50, and 10 µg/ml) to produce a volume of 1 ml. subsequently, the preparation was incubated for 20 minutes at 37°C using 2 mm ATP Tris salt to initiate the reaction. A 1 ml ice-cold solution of 10% v/v trichloroacetic acid was added to stop the reaction. Different dosages of the extract and omeprazole (100 µg/ml) were tested for their effects on the $H^+$-$K^+$ ATPase activity. Using spectrophotometry at 400 nm, the amount of inorganic phosphate liberated from ATP was recorded.

#### 7.2. Acid Neutralizing Capacity [14, 15]

The antacid, aluminum hydroxide + magnesium hydroxide (50 mg/ml), was compared with the acid neutralizing capacity of the sample mixture (EEJM) at concentrations of 500, 250, 100, 50, and 10 µg/ml. EEJM was diluted to 70 ml with water after adding 5 ml of the mixture and stirring for one minute. After adding 30 ml of 1.0 N HCl and stirring for 15 minutes, phenolphthalein drops were added. Excess HCl was titrated with 0.5 N NaOH until a pink color appeared.
Moles of acid neutralized were calculated as: Moles of acid neutralized = (Volume of HCl × Normality of HCl) - (Volume of NaOH × Normality of NaOH)

### 8. Invitro Anti-Inflammatory Activity
#### 8.1. Protein denaturation assay [16,17]

Following a modified method by Mizushima, Kobayashi, and Sakat et al., 500 µg of 1% bovine serum albumin was mixed with varying concentrations (500, 250, 100, 50, and 10 µg/ml) of EEJM test samples. After incubating at room temperature for 10 minutes, the mixture was heated at 51°C for 20 minutes and then cooled to room temperature. Absorbance at 660 nm was measured to assess protein denaturation. % Inhibition of protein denaturation was calculated using: % Inhibition = 100 - ((A1 - A2) / A0) * 100, where A1 is the absorbance of the control, A2 is the absorbance of the test sample, and A0 is the absorbance of the positive control. Triplicate experiments were conducted for accuracy.

### 8.2. Proteinase inhibitory assay [18]

The proteinase inhibitory experiment, based on the method by Oyedepo and Femurewa, involved preparing a reaction mixture containing 0.06 mg of trypsin, 1 ml of Tris-HCl buffer (20 mm, pH 7.4), and various concentrations of the test sample (EEJM) (500, 250, 100, 50, and 10 µg/ml). After incubating for five minutes at 37°C, 1 ml of 0.8% (w/v) casein was added and further incubated for twenty minutes. The reaction was terminated by adding 2 ml of 70% perchloric acid. Following centrifugation of the suspension, the absorbance of the supernatant was measured at 210 nm using a Tris-HCl blank. The experiment was conducted in triplicate to ensure reliability.

## 9. Results and Discussion

### 9.1. DPPH Method

DPPH is a stable free radical widely employed to assess antioxidant activity due to its characteristic absorbance at 517 nm, which decreases upon scavenging. Antioxidants reduce DPPH from purple to yellow, indicating their effectiveness. EEJM demonstrated concentration-dependent scavenging activity, with the highest inhibition at 500 µg/ml showing 88%. The IC50 value, a measure of potency, was calculated as 52.56 µg/ml, indicating EEJM's effective antioxidant capability compared to ascorbic acid used as a standard. The results were depicted in the table no.1 and figure 2.

**Table 1: DPPH radical scavenging activity of EEJM**

| S. No | Tested sample concentration (µg/ml) | Percentage | of inhibition | (in triplicates) | Mean value (%) |
|---|---|---|---|---|---|
| 1. | Ascorbic acid | 97.15762 | 96.64083 | 97.41602 | 97.07149 |
| 2. | 500 µg/ml | 75.71059 | 75.71059 | 75.96899 | 75.79673 |
| 3. | 250 µg/ml | 72.86822 | 71.05943 | 67.70026 | 70.54264 |
| 4. | 100 µg/ml | 65.63307 | 64.85788 | 62.0155 | 64.16882 |
| 5. | 50 µg/ml | 61.24031 | 60.98191 | 56.84755 | 59.68992 |
| 6. | 10 µg/ml | 39.01809 | 41.60207 | 38.75969 | 39.79328 |

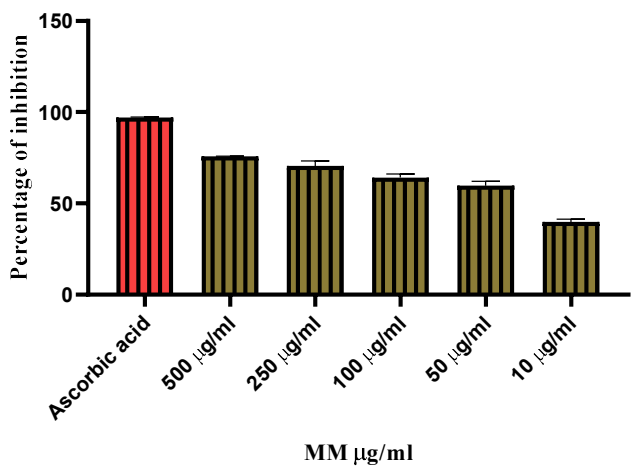

**Figure No.2: Percentage inhibition of EEJM by using DPPH**

## 9.2. ABTS Method

The ABTS radical cation assay measures antioxidant activity by monitoring absorbance at 734 nm, where a decrease indicates scavenging of the radical. Figure 3 illustrates the concentration-dependent ABTS radical scavenging activity of EEJM, with 500 µg/ml showing 75.7% inhibition. The IC50 value of EEJM was determined as 38.00 µg/ml, highlighting its potent antioxidant capability compared to standard antioxidants. This assay indicates EEJM's potential to protect vital tissues from oxidative damage. The results were depicted in the table no.2 and figure no.3.

**Table 2: ABTS radical scavenging activity of *Jatropha maheswari***

| S. No | Tested sample concentration (µg/ml) | Percentage of inhibition | (in triplicates) | | Mean value (%) |
|---|---|---|---|---|---|
| 1. | Ascorbic acid | 93.70079 | 91.33858 | 90.55118 | 91.86352 |
| 2. | 500 µg/ml | 88.34646 | 84.64567 | 91.10236 | 88.0315 |
| 3. | 250 µg/ml | 82.83465 | 83.54331 | 80.00 | 82.12598 |
| 4. | 100 µg/ml | 71.5748 | 72.59843 | 71.5748 | 71.91601 |
| 5. | 50 µg/ml | 67.6378 | 62.51969 | 62.20472 | 64.12073 |
| 6. | 10 µg/ml | 6.062992 | 18.18898 | 7.952756 | 10.73491 |

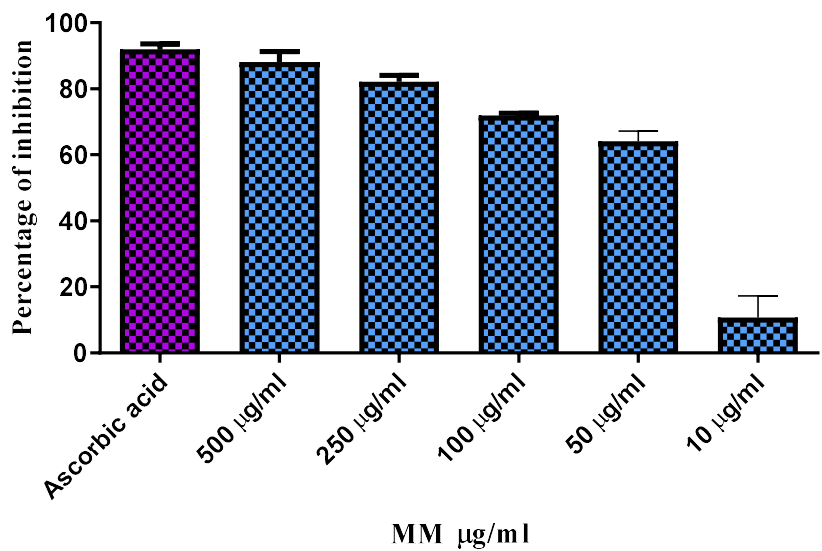

Figure No.3: Percentage inhibition of EEJM by using ABTS

### 9.3. In-Vitro Wound Healing Methods

#### 9.3.1. MTT Assay

The percentage of survived cells was calculated by measuring the absorbance of respective incubated cells in the 96 well s plate. The effect of the extracts on 3T3 cell lines is significant and comparable to the control. The extract has shown the activity even at the lowest concentration of 100μg/ml. The extract has shown concentration dependent activity. The extracts have shown the cell viability of 42.6% in the 1000 μg/ml concentration. This shows significant activity in MTT assay.

#### 9.3.2. Scratch Assay

In vitro wound scratch assays were performed using the 3T3 cell line. The healing abilities of ethanolic extracts were compared with a control. The extracts were tested for their ability to promote wound healing in 3T3 cells. A concentration of 10 mg/ml was chosen for the assays, which were run for 48 hours with wound closure measured at 24 and 48 hours. One extract showed significant wound healing activity, achieving a 98.27 closure at 48 hours. At 24 hours, the wound closure for this extract was 54%. The results of these assays are presented in the Tables 3 and fig 4 and 5.

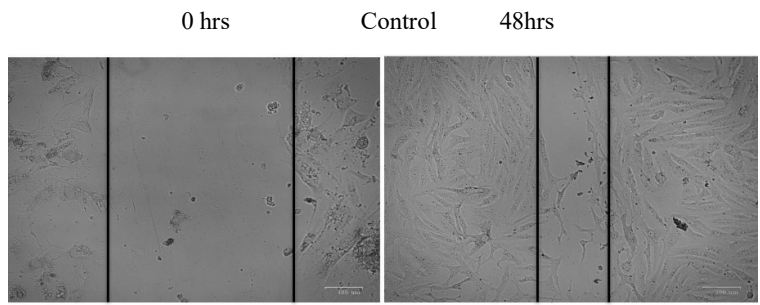

0 hrs     Control     48hrs

**Figure 4: Control at 48 Hrs**

0 hrs     TEST **(433 µg/ml of JM extract)**     48hrs

**Figure 4: Effect of JM at 48 Hrs**

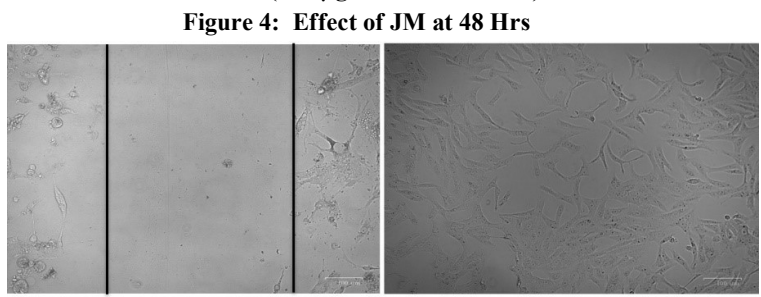

**Table 3: Densitometry analysis of wound healing assay by Image J software (24 hr)**

| Sample Details | Percentages of wound healing (in duplicates) | | Mean Value (%) |
|---|---|---|---|
| Control | 74.92 | 77.51 | 76.21 |
| Treated with 433 µg/ml MM sample | 98.67 | 97.88 | 98.27 |

## 9.4. In-Vitro Anti-Ulcer Activity

### 9.4.1. Assay of H+-K+-ATP ase

In the $H^+$-$K^+$-ATPase-induced ulcer assay, the ethanolic extract of *Jatropha maheswarii* demonstrated dose-dependent anti-ulcer effects when orally administered at 10 and 500 µg/ml, resulting in inhibition percentages of 10.6% and 41% ($p<0.001$), respectively, compared to the ulcer control. Omeprazole, a conventional medication, showed an 80.1% inhibition compared to the ulcer control. The IC50 value of the extract in the $H^+$-$K^+$-ATPase test was 76.94 µg/ml. The results are shown in Table 4 and Figure 6.

**Table 4: Percentage inhibition of EEJM**

| S.No | Tested sample concentration(µg/ml) | Percentage of Inhibition | IC 50 (µg/ml) |
|---|---|---|---|
| 1. | Omeprazole | 80.14403 | |
| 2. | EEJM 500 µg/ml | 41.04938 | |
| 3. | EEJM 250 µg/ml | 37.86008 | 76.94 |
| 4. | EEJM 100 µg/ml | 24.17695 | |
| 5. | EEJM 50 µg/ml | 25.00 | |
| 6. | EEJM 10 µg/ml | 10.69959 | |

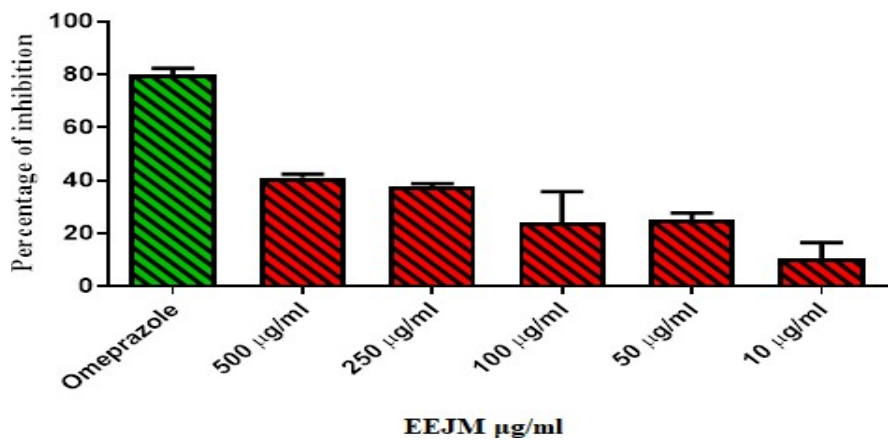

Figure 6: percentage inhibition of EEJM

### 9.4.2. Acid neutralizing capacity

The acid neutralizing capacity was used to test the ethanolic extract of *Jatropha maheswarii*. Oral administration of the extract at concentrations of 10 and 500 µg/ml of acid neutralizing capacity (ANC) per gram of antacid is 20 and 26 respectively. The results are shown in Table no 5 and Fig no 7.

Table: 5 Acid Neutralizing Capacity Of EEJM

| S.NO | Name of the sample | Name of the sample concentration | Reading a burette | Moles of acid neutralized | Acid neutralizing capacity (ANC) / antacid (g) |
|---|---|---|---|---|---|
| 1. |  | Control | 2.6 | 1.7 | 34 |
| 2. |  | 500 µg/ml | 3.4 | 1.3 | 26 |
| 3. |  | 250 µg/ml | 3.5 | 1.25 | 25 |
| 4. | EEJM | 100 µg/ml | 3.8 | 1.1 | 22 |
| 5. |  | 50 µg/ml | 3.9 | 1.05 | 21 |
| 6. |  | 10 µg/ml | 4.0 | 1.00 | 20 |

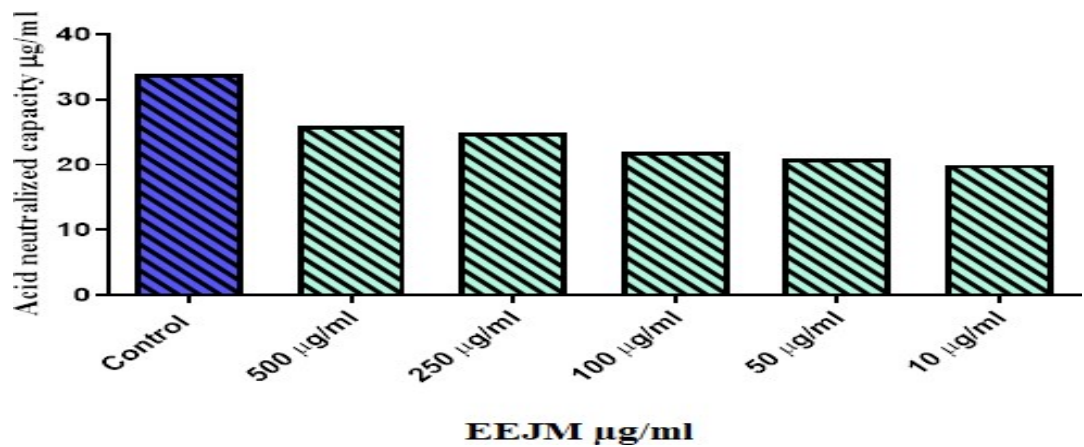

Figure 7: Acid neutralising

### 9.5. Invitro Anti-Inflammatory Activity
### 9.5.1. Protein Denaturation Assay

The current investigation demonstrated that the inhibition of albumin denaturation-induced inflammation was used to assess the *Jatropha maheswarii* ethanolic extract. The anti-inflammatory effect of oral administration of *Jatropha maheswarii* ethanol extract at concentrations of 10 and 500 µg/ml was demonstrated by dose-dependent inhibition percentages of 30.1 and 96.5% ($p<0.001$), respectively, as compared to the inflammation control. The percentage inhibition for the control medication was 100%. 135.0µg/ml is the IC50 value for the inhibition of albumin denaturation. The results are shown in Table 6 and Figure 8.

Table 6: Inhibition Percentage of Albumin Denaturation of EEJM

| S. No | Tested sample concentration(µg/ml) | Mean Value(%) | IC 50 (µg/ml) |
|---|---|---|---|
| 1. | Control | 100 | |
| 2. | EEJM 500 µg/ml | 96.55329 | |
| 3. | EEJM 250 µg/ml | 93.40136 | 135.0 |
| 4. | EEJM 100 µg/ml | 43.26531 | |
| 5. | EEJM 50 µg/ml | 36.73469 | |
| 6. | EEJM 10 µg/ml | 30.15873 | |

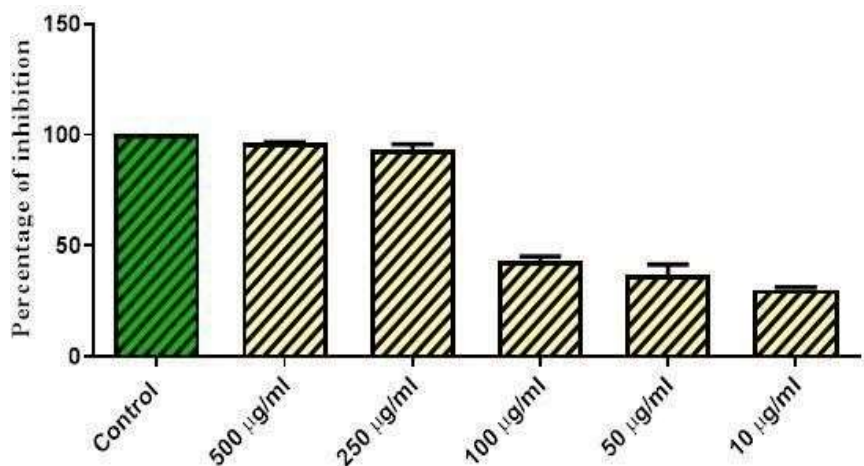
**Figure 8: Percentage of inhibition of albumin denaturation of EEJM**

### 9.5.2. Proteinase Inhibitory Assay

The current investigation demonstrated that the Proteinase Inhibitory Assay was used to assess the ethanolic extract of *Jatropha maheswarii* by inducing inflammation. The anti-inflammatory effect of oral administration of *Jatropha maheswarii* ethanol extract at concentrations of 10 and 500 µg/ml was demonstrated by dose-dependent inhibition percentages of 26.3 and 42.6% (p<0.001), respectively, as compared to the inflammation control. The percentage inhibition for the control medication was 100%. The Proteinase Inhibitory test has an IC50 value of 112.4 µg/ml. The results of these assays are presented in the Table 7.

**TABLE: 7 Proteinase Inhibitory assay of EEJM**

| S.No | Tested sample concentration(µg/ml) | Mean Value(%) | IC 50 (µg/ml) |
|---|---|---|---|
| 1. | Control | 100 | |
| 2. | EEJM 500 µg/ml | 42.66667 | |
| 3. | EEJM 250 µg/ml | 37.41176 | 112.4 |
| 4. | EEJM 100 µg/ml | 34.03922 | |
| 5. | EEJM 50 µg/ml | 30.43137 | |
| 6. | EEJM 10 µg/ml | 26.31373 | |

## 10. Conclusion:

The *invitro* assessment of antioxidant, wound healing, anti-ulcer and anti-inflammatory potential of ethanolic extract of *Jatropha maheswari* showed that the extract possesses significant activities. The extract was found to be effective in scavenging free radicals and inhibiting lipid peroxidation. The extract also promoted the proliferation of fibroblasts and the migration of keratinocytes. These results suggest that the ethanolic extract of *Jatropha maheswari* has the potential to be used as a therapeutic agent for the treatment of oxidative stress-related diseases and wounds. The extract has a substantial impact on the hydrogen potassium ATP area assay and acid neutralizing capacity. The plant extract's anti-inflammatory properties are demonstrated by a marked effect in the albumin denaturation assay's proteinase inhibition. It demonstrates that the extract habituated the prostaglandin and nitrous oxide production. Though many studies have been conducted, relatively few natural products have been commercialized or employed in clinical practice to aforesaid activities. In order to fully understand the potential of naturally occurring bioactive compounds further studies should be conducted to know the underlying mechanisms.